\documentclass[prl,aps,showpacs,twocolumn]{revtex4}

\usepackage{amsmath}
\usepackage{bm}
\usepackage{graphicx}

\begin{document}

\title{Split Instability of a Vortex in an Attractive Bose-Einstein
Condensate}

\author{Hiroki Saito}
\author{Masahito Ueda}
\affiliation{Department of Physics, Tokyo Institute of Technology,
Tokyo 152-8551, Japan}

\date{\today}

\begin{abstract}
An attractive Bose-Einstein condensate with a vortex splits into two
pieces via the quadrupole dynamical instability, which arises at a weaker
strength of interaction than the monopole and the dipole instabilities.
The split pieces subsequently unite to restore the original vortex or
collapse.
\end{abstract}


\maketitle

Quantized vortices in gaseous Bose-Einstein condensates (BECs) offer a
visible hallmark of superfluidity~\cite{Matthews,Madison,Abo}, where
repulsive interatomic interactions play a crucial role in the vortex
stabilization and lattice formation.
Attractive BECs, on the other hand, cannot hold vortices in any
thermodynamically stable state.
A fundamental issue of the decay of a many-particle quantum system may be
addressed if a vortex is created in an attractive BEC.
Such a state has become possible owing to the development of the Feshbach
technique~\cite{Inouye} by which the strength and the sign of
interactions can be controlled~\cite{Cornish,Roberts01,Donley}.

Suppose that a singly quantized vortex is created in a BEC with repulsive
interaction and that the interaction is adiabatically changed from
repulsive to attractive.
According to previous work~\cite{Dalfovo}, the vortex state remains
metastable until a dimensionless strength of interaction $g$ to be defined
later reaches a critical value $g_M^{\rm cr} (< 0)$, and when $|g|$ exceeds
$|g_M^{\rm cr}|$, the system develops a monopole (breathing-mode)
instability and collapses.
In this Letter, we show that this conclusion holds only when the system
has exact axisymmetry, and that even an infinitesimal symmetry-breaking
perturbation induces the quadrupole {\it dynamical instability} that
appears for $|g|$ smaller than $|g_M^{\rm cr}|$.
We note that similar dynamical instabilities~\cite{Recati,Sinha,Tsubota}
initiate vortex nucleation observed by the ENS group~\cite{Madison}.
A dynamical instability is also shown to transfer a vortex from one to the
other component of a binary BEC system~\cite{Garcia00}.
Here, we show that yet another dynamical instability causes a vortex to
split into two pieces that revolve around the center of the trap.
Surprisingly, in some parameter regimes, the pieces subsequently unite to
restore the original vortex, and this split-merge cycle repeats.
We report below the results of our studies on the collapsing dynamics of a
vortex in an attractive BEC.

We first investigate the Bogoliubov spectrum of a single-vortex state.
The single-vortex state is determined so as to minimize the
Gross-Pitaevskii (GP) energy functional within the axisymmetric functional
space $\psi_0 = f(r, z) e^{i\phi}$ with $r = (x^2 + y^2)^{1/2}$, where we
ignore the effect of vortex bending~\cite{Rosen,Garcia01,bending}.
In the following analysis, we normalize the length, time, and wave
functions in units of $d_0 \equiv (\hbar / m \omega_\perp)^{1/2}$,
$\omega_\perp^{-1}$, and $(N / d_0^3)^{1/2}$, where $\omega_\perp$ is the
radial trap frequency, and $N$, the number of BEC atoms.
We obtain the Bogoliubov spectrum by numerically
diagonalizing the Bogoliubov-de Gennes equations~\cite{Edwards}
\begin{subequations} \label{Bogo}
\begin{eqnarray}
\left( K + 2 g |\psi_0|^2 \right) u_n + g \psi_0^2 v_n & = & E_n u_n, \\
\left( K + 2 g |\psi_0|^2 \right) v_n + g \psi_0^{*2} u_n & = & -E_n
v_n,
\end{eqnarray}
\end{subequations}
where $K \equiv -\nabla^2 / 2 + (r^2 + \lambda^2 z^2) / 2 - \mu$ with
$\lambda \equiv \omega_z / \omega_\perp$, and $n$ is the index of the
eigenmode.
Here, $g \equiv 4\pi N a / d_0$ characterizes the strength of interaction,
where $a$ is the s-wave scattering length.
For a vortex state $\psi_0 \propto e^{i\phi}$, each angular momentum state
$u_n \propto e^{i m \phi}$ is coupled only to $v_n \propto e^{i (m - 2)
\phi}$, and we shall refer to $m$ as the angular momentum of the
excitation.

We find that there is at least one negative eigenvalue in the $m = 0$ mode
for any $g < 0$ and $\lambda$ even in the presence of a rotating drive.
The vortex state with attractive interactions is therefore
thermodynamically unstable, and eventually decays into the non-vortex
ground state by dissipating its energy and angular momentum.
At sufficiently low temperatures in a high-vacuum chamber, however, the
thermodynamic instability is irrelevant, since the energy and angular
momentum are conserved.
In fact, recent experiments~\cite{Madison} have demonstrated that the
vortex state in a stationary trap has a lifetime of $\sim 1$ s, which is
much longer than the characteristic time scales of the dynamics that we
shall discuss below.

When the complex eigenvalues emerge in the Bogoliubov spectrum, the
amplitude of the corresponding mode grows exponentially in time.
As noise is inevitable in experimental situations, such dynamical
instabilities are more important than the thermodynamic one at low
temperature.

\begin{figure}[tb]
\includegraphics[width=8.4cm]{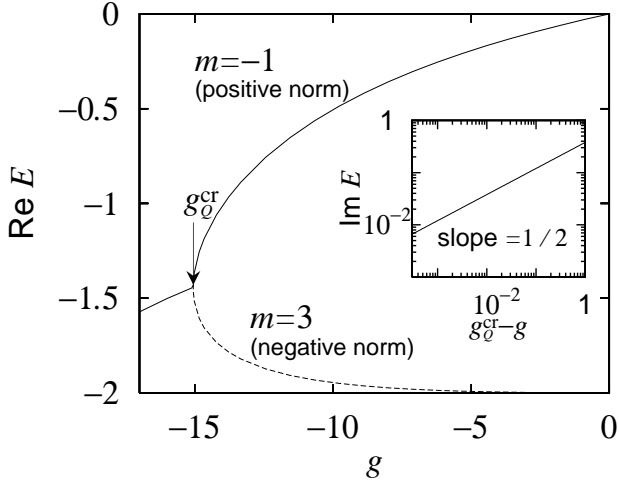}
\caption{
For $g > g_Q^{\rm cr} = -15.06$, the solid curve denotes the eigenvalue
(real) of the $m = -1$ mode, and the dashed one that of the negative-norm
branch of the $m = 3$ mode in an isotropic trap.
These two branches merge at $g = g_Q^{\rm cr}$, below which the eigenvalue
becomes complex and the dynamical instability sets in.
Inset: The solid line shows the imaginary part of the complex eigenvalues
for $g < g_Q^{\rm cr}$, which is proportional to $\sqrt{g_Q^{\rm cr} -
g}$.
}
\label{f:Bogo}
\end{figure}

Figure~\ref{f:Bogo} shows the real and imaginary parts of the lowest
eigenvalues of the $m = -1$ and $3$ excitations in an isotropic trap.
The eigenvalues become complex at the critical strength of interaction
$g_Q^{\rm cr} = -15.06$, showing the onset of the dynamical instability in
the quadrupole mode.
The imaginary part of the complex eigenvalue is proportional to
$\sqrt{g_Q^{\rm cr} - g}$ as shown in the inset in Fig.~\ref{f:Bogo}.
The complex eigenvalues emerge also in the dipole modes, i.e., $m = 0$ and
$2$, for $g < g_D^{\rm cr} = -18.02$.
The eigenvalues with other $m$ are real for $g$ larger than the critical
value for the monopole (radial-breathing-mode) instability $g_M^{\rm cr} =
-23.7$.

\begin{figure}[tb]
\includegraphics[width=8.4cm]{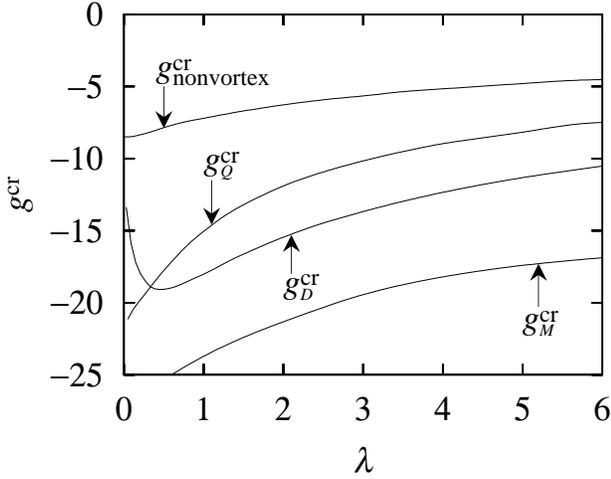}
\caption{
The $\lambda \equiv \omega_z / \omega_\perp$ dependence of the critical
strengths of interaction for the quadrupole mode $g_Q^{\rm cr}$, the dipole
mode $g_D^{\rm cr}$, and the monopole mode $g_M^{\rm cr}$ of the
single-vortex state and the critical strength of interaction for the
non-vortex ground state $g_{\rm nonvortex}^{\rm cr}$.
}
\label{f:lambda}
\end{figure}

Figure~\ref{f:lambda} shows the $\lambda$ dependence of $g_Q^{\rm cr}$,
$g_D^{\rm cr}$, $g_M^{\rm cr}$, and $g_{\rm nonvortex}^{\rm cr}$ in
axi-symmetric traps, where $g_{\rm nonvortex}^{\rm cr}$ is the critical
value for the non-vortex state to collapse through the monopole
instability.
We note that $|g_M^{\rm cr}|$ is always larger than $|g_Q^{\rm cr}|$ and
$|g_D^{\rm cr}|$, and hence the latter instabilities arise before the
monopole instability sets in.
For a trap with $\lambda \gtrsim 0.3$, the quadrupole instability arises
before the dipole one, unlike the quasi-1D toroidal
trap~\cite{Rokhsar,Ueda,Berman}.

We also performed numerical diagonalization for 2D systems.
Strong confinement in the $z$ direction produces the quasi-2D trap, when
$\hbar \omega_z$ is much larger than the characteristic energy of the
dynamics.
The effective strength of interactions in the quasi-2D oblate trap is
given by $g^{\rm 2D} = \sqrt{\lambda / (2\pi)}g^{\rm 3D}$~\cite{Castin}.
In the 2D system, the dynamical instability arises in the quadrupole,
dipole, and monopole modes at $g_Q^{\rm 2D cr} = -7.79$, $g_D^{\rm 2D cr}
= -11.48$, and $g_M^{\rm 2D cr} = -23.4$.
The dependencies of the complex eigenvalues on $g$ are similar to that in
Fig.~\ref{f:Bogo}.

To understand the dynamical instabilities analytically, let us consider
the GP action integral in 2D
\begin{equation} \label{action}
S = \int dt \int d{\bf r} \psi^* \left( -i \frac{\partial}{\partial t} -
\frac{\nabla^2}{2} + \frac{r^2}{2} + \frac{g}{2} |\psi|^2 \right) \psi.
\end{equation}
We assume that the state evolution is described by $\psi = \sum_m c_m(t)
\phi_m({\bf r})$, where $\phi_m({\bf r})$ is assumed to take the form of
$\phi_m({\bf r}) = [r^{|m|} / (\sqrt{\pi |m|!} d^{|m| + 1})] \exp[-r^2 /
(2d^2) + i m \phi]$ and $d = [1 + g / (8\pi)]^{1/4}$ minimizes the GP
energy functional for the $m = 1$ state.
Substituting this $\psi$ into Eq.~(\ref{action}) and minimizing $S$ with
respect to $c_m$ yields
\begin{equation} \label{cdot}
i \dot c_m = \varepsilon_m c_m + g \sum_{m m_1 m_2 m_3} G_{m_2, m_3}^{m,
m_1} c_{m_1}^* c_{m_2} c_{m_3},
\end{equation}
where $\varepsilon_m \equiv \int d{\bf r} (|\nabla \psi_m|^2 + r^2
|\psi_m|^2) / 2$ and $G^{m_1, m_2}_{m_3, m_4} \equiv \int d{\bf r}
\psi_{m_1}^* \psi_{m_2}^* \psi_{m_3} \psi_{m_4}$.
When the BEC exists in the $m = 1$ mode, we obtain $c_1(t) = e^{-i \mu t}
+ O(|c_{m \neq 1}|^2)$ with $\mu = 1 / d^2 + d^2 + g / (4 \pi d^2)$.
The linear analysis of Eq.~(\ref{cdot}) for $\tilde c_m \equiv e^{i \mu t}
c_m$ $(m \neq 1)$ yields
\begin{equation}
i \dot{\tilde c}_m = (\varepsilon_m - \mu) \tilde c_m + 2 G^{m, 1}_{m, 1}
\tilde c_m + G^{m, 2 - m}_{1, 1} \tilde c_{2-m}^*.
\end{equation}
It follows from this that, for $m = -1$, the eigenfrequencies are given by
$A \pm \sqrt{B}$, where $A \equiv [g / (8\pi) - 1] / [1 + g /
(8\pi)]^{1/2}$ and $B \equiv 3 + 5g / (8\pi) + [1 + g / (2\pi) - g^2 /
(32\pi^2)] / [1 + g / (8\pi)]$.
We find that $B$ is a monotonically increasing function for $g > -8\pi$,
and $B$ becomes negative for $g < g^{\rm cr} \simeq -9.2$, which is in
reasonable agreement with $g_Q^{\rm 2D cr} = -7.79$ stated above.
We also find that the imaginary part appearing for $g < g^{\rm cr}$ is
proportional to $\sqrt{g^{\rm cr} - g}$, in agreement with the inset of
Fig.~\ref{f:Bogo}.

The Bogoliubov analysis described above is valid only if deviations from
a stationary state are small.
To follow further evolution of the wave function, we must solve the
time-dependent GP equation.
Since we are studying the growth of small perturbations, high precision is
required in the numerical integration, and hence we consider the GP
equation in 2D
\begin{equation} \label{GP}
i \frac{\partial \psi}{\partial t} = \left[ -\frac{1}{2} \nabla^2 +
\frac{1}{2} r^2 + g^{\rm 2D} |\psi|^2 \right] \psi
\end{equation}
to ensure sufficiently small discretization in the Crank-Nicholson
scheme~\cite{Ruprecht}.
This situation corresponds to an oblate trap with large $\lambda$.

\begin{figure}[tb]
\includegraphics[width=8.2cm]{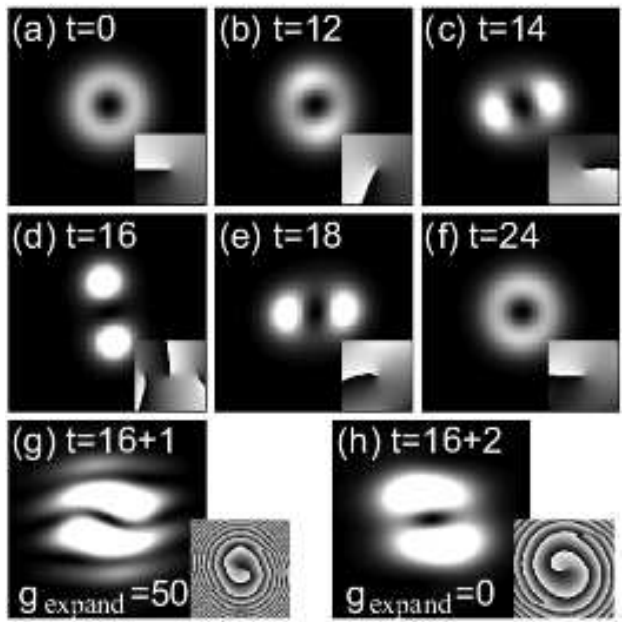}
\caption{
The density and phase (insets) profiles of the time evolution of the
vortex state.
The initial state is the stationary solution of the Gross-Pitaevskii
equation plus a small symmetry-breaking perturbation.
The insets present the gray-scale plots of the phase from $(2 n - 1) \pi$
to $(2 n + 1) \pi$ with an integer $n$.
(a)-(f) Time evolution with $g^{\rm 2D} = -9$.
(g) An expanded image at $t = 17$, where the interaction is switched from
$g^{\rm 2D} = -9$ to $g_{\rm expand} = 50$, and the trap is switched off
at $t = 16$.
(h) An expanded image at $t = 18$ with $g_{\rm expand} = 0$.
The sizes of the images are $7 \times 7$ in (a)-(f) and $18 \times 18$
in (g) and (h) in units of $(\hbar/m\omega_\perp)^{1/2}$.
The sensitivity of imaging in (g) and (h) is 20 times higher than that in
(a)-(f).
}
\label{f:split}
\end{figure}

Figures \ref{f:split} (a)-(f) depict the time evolution of the density and
phase profiles with $g^{\rm 2D} = -9$, which is smaller than the critical
value for the quadrupole mode $g_Q^{\rm 2D cr} = -7.79$ but larger than
that for the dipole mode $g_D^{\rm 2D cr} = -11.48$.
A small symmetry-breaking perturbation is added to the initial state to
imitate noise in realistic situations.
Due to the quadrupole instability, the vortex is first stretched
[Fig.~\ref{f:split} (b)], and then splits into two clusters that revolve
around the center of the trap [Fig.~\ref{f:split} (d)] with angular
velocity $\simeq 0.73 \omega_\perp$.
In the first deformation process the $m = -1$ and $3$ components grow
exponentially, and their Lyapunov exponents agree with the imaginary part
of the complex eigenvalues.
Interestingly, the split process is reversible: the two clusters
subsequently unite to restore the ring shape [Fig.~\ref{f:split} (f)], and
this split-merge process repeats.
We numerically checked that no split-merge phenomenon occurs for $g^{\rm
2D} > g_Q^{\rm 2D cr}$, where the system is metastable.
The insets in Fig.~\ref{f:split} illustrate the phase plots.
At $t = 16$, there are three topological defects: the central one exists
from the outset, and the other two enter as the vortex splits, in
accordance with the fact that the $m = 3$ component grows upon the vortex
split.
The two side vortices cannot be seen in the density plot, and hence they
may be called ``ghost'' vortices~\cite{Tsubota} that carry very little
angular momentum.

The two clusters in Fig.~\ref{f:split} (d) may be regarded as revolving
``solitons'' whose phases differ by $\pi$.
In fact, at an energy only slightly below that of Fig.~\ref{f:split} (d),
there is a low-lying state in which two solitons revolve without
changing their shapes.
It is interesting to note that this situation is similar to the
soliton-train formation observed by the Rice group~\cite{Strecker}, where
the modulation (dynamical) instability causes a quasi-1D condensate to
split into solitons when the interaction is changed from repulsive to
attractive using the Feshbach resonance.
We note that similar instabilities split an optical vortex propagating in
a nonlinear medium into spiraling
solitons~\cite{Tikhonenko,Garcia_opt,Mihalache}.
This similarity between attractive BECs and optical
solitons~\cite{Stegeman} implies that other nonlinear phenomena, such as
pattern formation, which has been predicted in attractive
BECs~\cite{Saito}, may also be realized in optical systems.

When the system becomes too small to be observed by the {\it in situ}
imaging method due to the attractive interaction, the condensate must be
expanded before imaging.
Figure~\ref{f:split} (g) shows the expanded image at $t = 17$, where the
interaction is switched from $g^{\rm 2D} = -9$ to $g_{\rm expand} = 50$
and the trapping potential is switched off at $t = 16$ [Fig.~\ref{f:split}
(d)].
The image shows the interference fringes due to the overlap of the atomic
clouds emanating from the two clusters.
The wavelength of the interference pattern is proportional to the
expansion time.
Figure~\ref{f:split} (h) shows the expanded image at $t = 18$ with
$g_{\rm expand} = 0$.
Comparing Figs.~\ref{f:split} (g) and (h), we find that the stronger
repulsive interaction produces more fringes and bends them around the
center.

When $|g^{\rm 2D}|$ exceeds $|g_D^{\rm 2D cr}| = 11.48$, a dipole
instability arises in addition to the quadrupole one.
The dipole instability causes atoms to transfer from one cluster to the
other, thereby inducing the collapse.
\begin{figure}[tb]
\includegraphics[width=8.2cm]{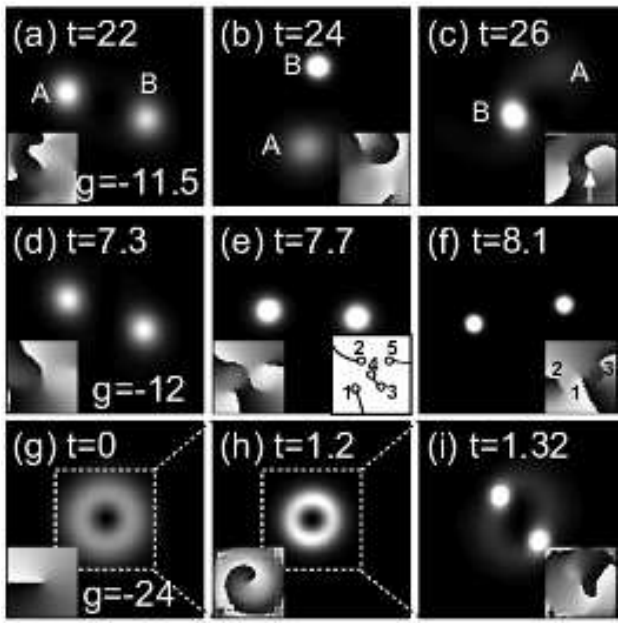}
\caption{
The density and phase (insets) profiles of the time evolution after the
vortex split with $g^{\rm 2D} = -11.5$ in (a)-(c) and with $g^{\rm 2D} =
-12$ in (d)-(f).
In (g)-(i), $g^{\rm 2D}$ is switched from $0$ to $-24$ at $t = 0$.
The right inset in (e) illustrates the topological defects and branch cuts
in the left one.
Here we note that the end part of the central branch cut is separated to
yield a vortex-antivortex pair 3-4 [inset of (c)].
Vortex 5 and antivortex 4 are subsequently combined [inset of
(f)].
The sizes of the images are $5 \times 5$ in (a)-(f), $8 \times 8$ in (g),
$4 \times 4$ in (h), and $2 \times 2$ in (i) in units of
$(\hbar/m\omega_\perp)^{1/2}$.
The sensitivity of imaging of the density plots is $1 / 4$ in (a)-(f), $1$
in (g), $1 / 5$ in (h), and $1 / 20$ in (i) in units of those in
Figs.~\ref{f:split} (a)-(f).
}
\label{f:collapse}
\end{figure}
Figures~\ref{f:collapse} (a)-(c) show the collapse process with $g^{\rm
2D} = -11.5$.
After the split-merge process repeats a few times, the balance between
the two clusters is broken due to the dipole instability.
As a consequence, the cluster labeled A grows [Fig.~\ref{f:collapse} (a)],
then B grows [Fig.~\ref{f:collapse} (b)] like a seesaw, and eventually B
absorbs most atoms and collapses [Fig.~\ref{f:collapse} (c)], where the
original topological defect begins to spiral out as indicated by the white
arrow in the inset.
With a stronger attractive interaction $g^{\rm 2D} = -12$, both clusters
collapse immediately after the vortex split as shown in
Figs.~\ref{f:collapse} (d)-(f).
In this collapse process, we found the exchange of a vortex-antivortex
pair (see the insets).
This phenomenon is also seen in the split-merge process with $g^{\rm 2D} =
-11.5$, while it is not seen at weaker attractive interactions, say, at
$g^{\rm 2D} = -9$.

Figures~\ref{f:collapse} (g)-(i) show the collapse where the interaction
is switched from $g^{\rm 2D} = 0$ to $g^{\rm 2D} = -24 < g_M^{\rm 2D cr}$
at $t = 0$.
The vortex first shrinks due to the monopole instability [(g)
$\rightarrow$ (h)], then splits into two clusters due to the quadrupole
instability [(h) $\rightarrow$ (i)], and both clusters collapse.
When the interaction is switched to an even greater attractive one, a
shell structure is formed~\cite{Saito}, which splits into several parts
due to multipole instabilities, and each fragment collapses and
explodes~\cite{Donley}, producing very complicated collapsing dynamics.

In conclusion, we have studied the dynamical instabilities of a quantized
vortex in an attractive BEC.
The dynamical quadrupole instability spontaneously breaks the axisymmetry
and splits a vortex into clusters that revolve around the center of the
trap, which then unite to restore the vortex or eventually collapse.
The dynamical instabilities presented here play a much larger role than
the thermodynamic one at low temperature, and serve as a dominant
mechanism for the collapsing dynamics of a rotating condensate: vortices
collapse via the dynamical instabilities around the topological defects.

This work was supported by a Grant-in-Aid for Scientific Research (Grant
No. 11216204) by the Ministry of Education, Science, Sports, and Culture
of Japan, and by the Toray Science Foundation.


\begin{thebibliography}{}
\bibitem{bending}
In recent experiments~\cite{Rosen}, the time scale of the vortex bending
is found to be $\gtrsim$ 1 s, which is much larger than the time scale of
the vortex split ($\sim$ 10 ms).
This is because the former is due to a thermally activated
process~\cite{Rosen}, and the latter, to a dynamical instability.
Furthermore, the bending effect will be much less pronounced in a
spherical or prolate trap, which we assume in this Letter, than in an
elongated trap.
\end{thebibliography}
\end{document}